# The stability and physical properties of the tetragonal phase of bulk CuMnAs antiferromagnet

Klára Uhlířová, Elen Duverger-Nédellec, Ross Colman, Jiří Volný, Barbora Vondráčková, Karel Carva

*Department of Condensed Matter Physics, Faculty of Mathematics and Physics, Charles University in Prague, Ke Karlovu 5, 12116 Prague, Czech Republic*

*Abstract*

The effect of Cu substitution on the stability of the CuMnAs tetragonal phase was studied both experimentally and by *ab initio* calculations. Polycrystalline samples with various compositions $Cu_{1+x}Mn_{1-x}As$ (x = 0-0.5) were synthetized. The tetragonal phase of CuMnAs is found to be stabilized by substituting Mn by Cu in the amount of $x \sim 0.1$ or higher. This observation is supported by *ab initio* calculations of the total energy of the tetragonal and orthorhombic phases; with increasing Cu content the tetragonal phase is favoured. Small variations of composition thus allow to grow selectively one of these two phases with distinct and unique features for antiferromagnetic spintronics.

Measurements of magnetic susceptibility and differential scanning calorimetry have shown that the tetragonal $Cu_{1+x}Mn_{1-x}As$ has an antiferromagnetic behaviour with the maximum Néel temperature $T_N$ = 507 K for the highest Mn content samples, decreasing with the decreasing Mn content.

## 1. Introduction

The antiferromagnetic semi-metal CuMnAs, which exists in both tetragonal and orthorhombic structure, have been attracting recent attention of experimental and theoretical physicists in the fields of antiferromagnetic spintronic and physics of Dirac fermions [1–3]. A controlled rotation of magnetic moments' orientation by means of an applied electrical field has been demonstrated in tetragonal CuMnAs [4]. This reorientation is achieved by the creation of staggered fields due to the spin-orbit torques in antiferromagnets of specific symmetry [5]. Such effect allows for creation of a unique memory device and paves the way for antiferromagnetic spintronics [6–8]. On the other hand, the orthorhombic CuMnAs has been proposed to be the first candidate for a Dirac semimetal with magnetic order [3]. It is thus desirable to describe the conditions that lead to the formation of the orthorhombic and tetragonal phases.

The crystal structure of the compounds from the ternary Cu-Mn-As system is sensitive to the composition. Three different structures have been reported for this system:

(i) The stoichiometric or almost stoichiometric compound crystallizes in an orthorhombic structure with space group Pnma (Figure 1 b) and with lattice parameters $a$ = 0.6586 nm, $b$ = 0.3867 nm, $c$ = 0.7320 nm [8–10].

(ii) slightly As rich (7%) and/or Cu rich (3%) and Mn deficient samples form a tetragonal structure with space group P4/nmm (Figure 1 a) and with lattice parameters $a$ = 0.3800(4) nm, $b$ = 0.6328(10) nm [11]. Similar lattice parameters $a$ = 0.3820(10) nm, $b$ = 0.6318(10) nm were reported for CuMnAs thin films grown on GaAs (001) substrate [12,13]. Recent studies [14] suggest that the tetragonal thin films grown on GaAs (001) substrate are also Mn deficient.

(iii) when going further to the Mn-rich side of the ternary diagram, namely $Cu_2Mn_4As_3$ (= $Cu_{0.67}Mn_{1.33}As$) and $CuMn_3As_2$ (= $Cu_{0.50}Mn_{1.5}As$) an orthorhombic II structure (Figure 1c) with doubled unit cell along the *a* axis compared to stoichiometric orthorhombic CuMnAs is formed [15,16].

Both the orthorhombic and tetragonal phases are reported to be room temperature antiferromagnets [8,9,13,17]. The Néel temperature $T_N$ of the orthorhombic CuMnAs ranges between 330 – 360 K [8,10], a more complex modulated structure was recently reported in off-stoichiometric orthorhombic $Cu_{0.98}Mn_{0.96}As$ single crystals [10]. The magnetic structure of the tetragonal phase has only been studied in epitaxial thin films, the Néel temperature reaches (480 ±5) K [17,18] and is therefore significantly higher than in the orthorhombic phase. *Ab initio* calculations have been used to examine both the orthorhombic [8] and the tetragonal phases [13,19]. The Neel temperature obtained by atomistic spin dynamics from the calculated exchange interactions corresponds very well to the experiment in the latter case [19].

In this paper, we aim to describe the synthesis of bulk tetragonal CuMnAs prepared by solid-state reaction as well as the stability of the tetragonal phase for various Mn:Cu ratios from both experimental and a density functional theory perspective. The Néel temperature dependence of the Mn concentration was studied.

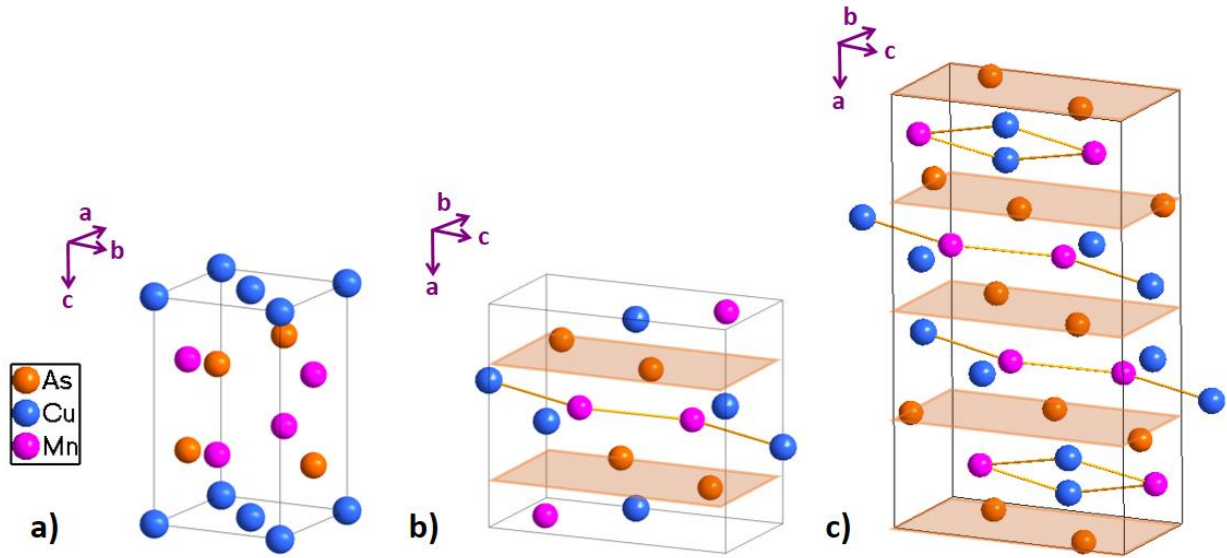

Figure 1. The tetragonal (a), orthorhombic (b) and orthorhombic II (c) structure of $Cu_{1+x}Mn_{1-x}As$ compounds. The shaded planes are a guide to the eye, highlighting similarities between orthorhombic cells.

## 2. Experimental and computational details:

The polycrystalline samples were prepared by a reaction of high purity Cu, Mn and As. Small pieces of the elements with desired molar ratios and total weigh of 3-4 g were placed in high purity alumina crucibles and sealed under 0.2 bar of Ar atmosphere in fused quartz ampoules. Each sample was then slowly (0.1 °C/min) heated up to 1090 °C where it was kept for 24 h, then it was cooled to 980 °C and held for 7 days to improve homogeneity. These temperatures are above the melting points of MnAs (935 °C) and CuMnAs (950°C [9]). After the synthesis, only very small amounts of a white film was found at the walls of the ampoules suggest that negligible amount of As sublimated from the sample. The synthesis resulted in shiny polycrystals with relatively large grains of the material.

The composition analysis was performed using scanning electron microscope (SEM) Tescan Mira LMH equipped with an energy dispersive x-ray detector (EDX) based on a non-standard method with precision up to 1-2%. The crystal structure was determined both by powder and single-crystal x-ray diffraction (XRD). Powder XRD data were collected on Bruker D8 Advance diffractometer in standard Bragg-Brentano geometry. Lattice parameters were determined by the LeBail method using Fullprof [20]. The single crystal data were collected at room temperature using a four-circle diffractometer Gemini (Agilent) with kappa geometry, equipped with a Mo sealed X-Rays tube with graphite collimator and with a CCD Atlas detector. Data reduction was made using CrysAlisPro software. The structure resolution and the refinement were done using Jana2006 software [21].

Polycrystalline samples were subject to differential scanning calorimetry DSC using Setaram Setsys calorimeter in the temperature range 300-600 K and magnetization measurements in temperature the range 300-600 K using a Quantum Design Physical Property Measurement System with the VSM Oven option. Higher temperatures were not used to minimize a risk of As sublimation and subsequent contamination of the equipment.

Calculations were based on the density functional theory employing the Green function tight-binding linear muffintin orbital method [22] and the atomic–sphere approximation (TB-LMTO-ASA) [23]. Moreover, the local spin-density approximation, the Vosko-Wilk-Nursair exchange potential [24], and the *s, p, d* basis are used. We have included relativistic corrections by means of adding the on-site spin-orbit coupling term to the scalar-relativistic TB-LMTO Hamiltonian. Small concentrations of dopants in some of the samples render the coherent potential approximation (CPA) [25,26] to be highly efficient here. The atomic basis contains two formula units (6 atoms) in the case of tetragonal structure and four formula units in the orthorhombic cases. More details on the construction of elementary cell can be found in Ref. 19. For the orthorhombic phase we consider the energetically most favourable magnetic ordering according to the previous study (↑↑↓↓) [3].

## 3. Results and discussion

*Crystal structure*

Samples with compositions $Cu_{1+x}Mn_{1-x}As_{1+y}$ with a small variation of As starting composition (to compensate sublimation) were prepared. The prepared materials were brittle with relatively large grains (submillimetre size). From each polycrystal we isolated a smaller piece for powder XRD and single crystal grains for single crystal XRD. Prior to the XRD measurements the pieces were analysed by SEM EDX. Depending on the composition, the samples had orthorhombic or tetragonal structure. The typical XRD of the tetragonal phases are shown in Figure 2.

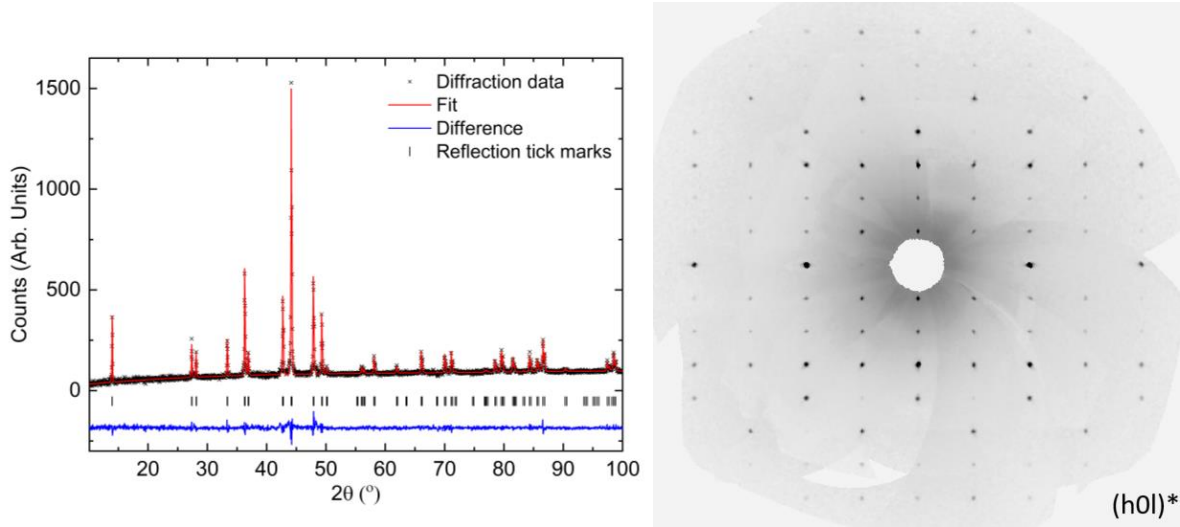

Figure 2. PXRD (a) and single crystal XRD (b) pattern of the tetragonal phase. The single crystal data were obtained on a sample with dimension 50x20x5 um$^3$.

The crystals structure, lattice parameters obtained by PXRD, composition determined by the EDX analysis and starting growth compositions are summarized in Table 1. For completeness of the $Cu_{1+x}Mn_{1-x}As$ system, we also include data from literature. In Figure 3, the relevant compounds of the Cu-Mn-As ternary system including crystal structure are presented in a ternary diagram; the sample compositions are coded in colours, and the structure is coded by symbol shape. We can see that the stoichiometric CuMnAs crystallizes in the orthorhombic structure as reported before [9], however, the border between the tetragonal and orthorhombic phase is very close to the stoichiometric CuMnAs composition. Our EDX analysis has shown that the standard deviation of composition at different spots within one sample is larger ($\pm 0.02$-0.04) than average the difference in compositions between the closest tetragonal and orthorhombic phases.

By focusing on the substitution of Mn by Cu and keeping As content constant, i.e. varying the composition along the $Cu_{1+x}Mn_{1-x}As$ line ($x = 0 - 0.51$), we found the border between the orthorhombic and tetragonal phase being between nominal (starting) compositions $Cu_{1.061}Mn_{0.939}As_{1.000}$ (EDX composition $Cu_{1.06(2)}Mn_{0.98(2)}As_{0.96}$) and $Cu_{1.152}Mn_{0.848}As_{1.000}$ (EDX composition $Cu_{1.15(2)}Mn_{0.87(3)}As_{0.97(3)}$). The most Cu-rich compound we prepared was with the nominal composition $Cu_{1.515}Mn_{0.455}As_{1.000}$. According to SEM EDX this sample contained majority of $Cu_{1.44(2)}Mn_{0.60(3)}As_{0.98(3)}$ and small amount of $Cu_{2.90(2)}Mn_{0.10(5)}As_{0.98(3)}$.

Considering some As evaporation, we prepared several samples with slight excess of As. It resulted in a material with the closest composition to stoichiometric CuMnAs which already had the tetragonal structure. This material was prepared from nominal composition $Cu_{1.00}Mn_{0.964}As_{1.036}$ and the composition from EDX analysis was determined as $Cu_{1.02(1)}Mn_{0.99(2)}As_{0.99(2)}$.

From this observation we may conclude that Cu excess stabilizes the tetragonal phase, however, the arsenic content, especially in the central part of the ternary system, also plays an important role. When the starting composition has a slight (~3%) excess of As, the tetragonal phase is more stable (is formed with higher Mn and lower Cu content). However, it is difficult to specify the composition of the samples precisely. Although the EDX analysis shows systematically lower (~3%) arsenic content compare to starting composition, it is not precise enough to be conclusive. On the other hand, we cannot simply work on the assumption that the final and starting composition is the same because part of the arsenic excess sublimates, and part of it can form a MnAs binary phase; both effects are difficult to quantify because we can never analyse the homogeneity of the whole sample from the microscopic point of view. Sublimation of As was negligible in most of the batches (only thin white coverage was found at the surface of the quartz glass) except the two most Cu-rich samples, where As crystals were found at the walls of the quartz glass ampoule in the amount almost compensating the excess of the element. Formation of the MnAs binary phase was detected in very small amount in one of our batches (nominal composition $Cu_{1.00}Mn_{0.964}As_{1.036}$) by susceptibly measurements, where a ferromagnetic transition ~315 K is evident although it was not found by SEM EDX analysis. MnAs was also reported by Emmanoulidou et al., [10] in samples containing a 7% arsenic excess.

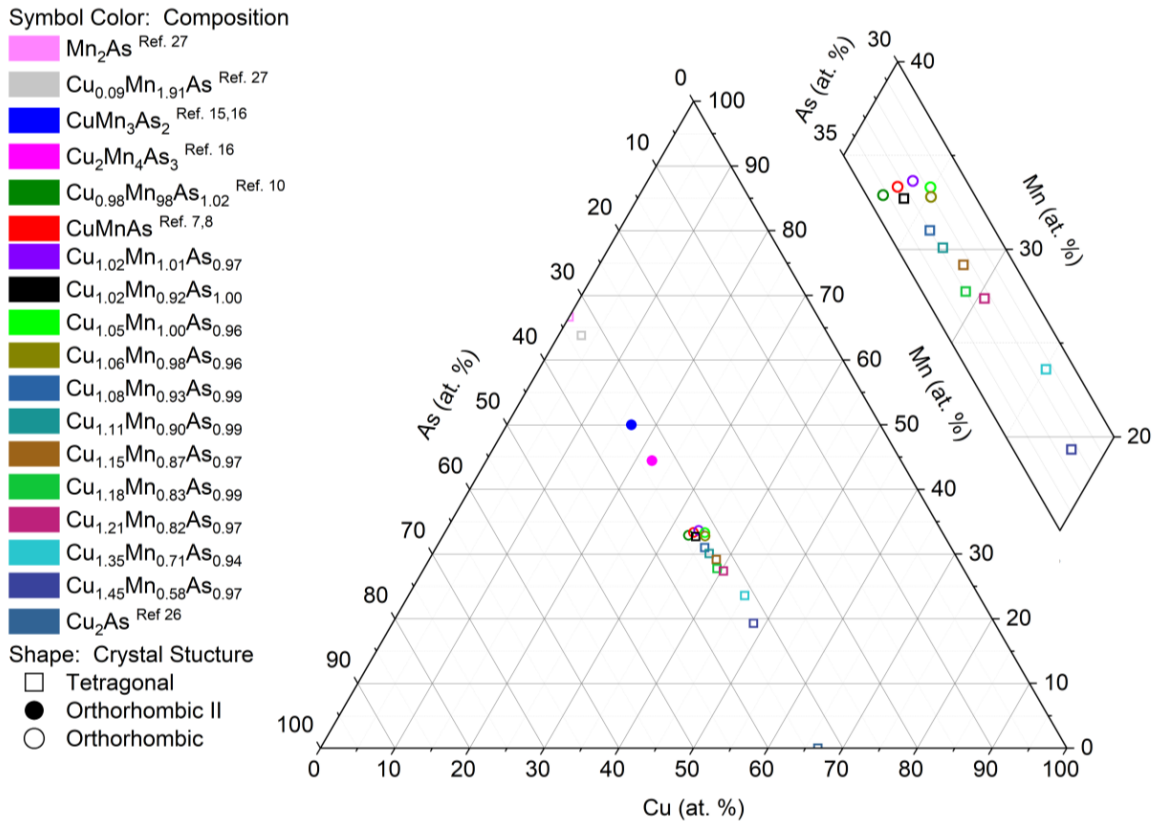

Figure *3*. Cu-Mn-As ternary diagram. The colour coding refers to different compositions, the tetragonal phase is presented by symbol □, the orthorhombic phase by symbol ○ and the orthorhombic II phase by symbol ●.

Table 1-The starting composition of prepared samples, composition determined by SEM EDX and crystal structure of the resulting compounds.

| Starting composition | EDX composition | Space group | Lattice parameters (nm) | Ref. |
|---|---|---|---|---|
| $Mn_2As$ | $Mn_2As$ | P4/nmm | a = 0.3769, c = 0.6278 | [27] |
| $Cu_{0.09}Mn_{1.91}As$ | $Cu_{0.09}Mn_{1.91}As$ | P4/nmm | - | [27] |
| $CuMn_3As_2$, Bi-flux | $CuMn_3As_2$ | Pnma | a = 1.3067(3), b = 0.38170 (8) c = 0.73190 (15) | [15,16] |
| Bi-flux | $Cu_2Mn_4As_3$ | Pnma | a = 1.3067(3), b = 0.38170 (8) c = 0.73170 (15) | [16] |
| Bi-flux | $Cu_{0.95}MnAs$ | Pnma | a = 0.65716(4), b = 0.38605(2) c = 0.73047(4) | [10] |
| CuMnAs | CuMnAs | Pnma | a = 0.65913, b = 0.38657 c = 0.73036 | [8,9] |
| $Cu_{1.000}Mn_{0.964}As_{1.036}$ | $Cu_{1.02}Mn_{0.92}As_{1.00}$ | Pnma | a = b = 0.37971(6), c = 0.6346(11) | |
| $Cu_{1.100}Mn_{0.900}As_{1.000}$ | $Cu_{1.05}Mn_{1.00}As_{0.96}$ | P4/nmm | a = 0.6601(2), b = 0.3859(2) c = 0.7315(3) | |
| $Cu_{1.061}Mn_{0.939}As_{1.000}$ | $Cu_{1.06}Mn_{0.98}As_{0.96}$ | Pnma | a = 0.6606(3), b = 0.3861(2), c = 0.7316 | |
| Bi-flux | $Cu_{1.08}Mn_{0.93}As_{0.99}$ | P4/nmm | a = b = 0.3808(7), c = 0.6345(8) | |
| $Cu_{1.082}Mn_{0.899}As_{1.020}$ | $Cu_{1.11}Mn_{0.90}As_{0.99}$ | P4/nmm | a = b = 0.38086(19), c = 0.63342(3) | |
| $Cu_{1.140}Mn_{0.840}As_{1.000}$ | $Cu_{1.15}Mn_{0.87}As_{0.97}$ | P4/nmm | a = b = 0.38154(4), c = 0.6320(10) | |
| $Cu_{1.182}Mn_{0.799}As_{1.020}$ | $Cu_{1.18}Mn_{0.83}As_{0.99}$ | P4/nmm | a = b = 0.38057(2), c = 0.63081(4) | |
| $Cu_{1.200}n_{0.780}As_{0.990}$ | $Cu_{1.21}Mn_{0.58}As_{0.97}$ | P4/nmm | a = b = 0.38064(6), c = 0.6316(15) | |
| $Cu_{1.515}Mn_{0.650}As_{1.020}$ | $Cu_{1.35}Mn_{0.71}As_{0.94}$ | P4/nmm | a = 0.38154(14), c = 0.62805 (3) | |
| $Cu_{1.515}Mn_{0.455}As_{1.000}$ | $Cu_{1.45}Mn_{0.58}As_{0.97}$ | P4/nmm | a = 0.38069(18), c = 0.6255(3) | |
| $Cu_2As$ | $Cu_2As$ | P4/nmm | a = 0.3788, c = 0.5942 | [27] |

*The data from literature include either starting composition or composition from EDX analyses depending on which information was available.*

In further text we will refer to compositions determined by EDX. The value will always be including the indicated error in brackets (while nominal compositions determined with much higher precision is without indicated errors)

*Calculations of the structure stability*

In order to examine further the conditions stabilizing the tetragonal phase, we have performed *ab initio* calculations where we varied the Cu concentration on the Mn sublattice. The total energy difference between the tetragonal and orthorhombic phase $E_{tet}$ - $E_{ort}$ as a function of the amount of extra Cu atoms (1 + x) on Mn sublattice is plotted in Figure 4, showing a clear tendency towards the tetragonal phase with increasing Cu content. In the calculation we have held lattice parameters constant, we have used here the values of samples (with composition $Cu_{1.02(1)}Mn_{0.99(2)}As_{0.99(2)}$) that are most close to the transition, which corresponds to *c/a* ratio 1.67 and Wigner-Seitz radius $r_{WS}$ = 2.90 $a_0$, where $a_0$ denotes the Bohr radius. We should note that the total energy depends strongly on lattice parameters, and a numerical optimization of lattice parameters for different substituent concentrations would improve further the general understanding of this system. This, however, goes beyond the scope of the method we had employed. The critical Cu concentration where the phase transition occurs is sensitive to a number of parameters and its real value can differ from the calculated one. However, the overall energetic preference for the tetragonal phase with increasing Cu content is a robust feature.

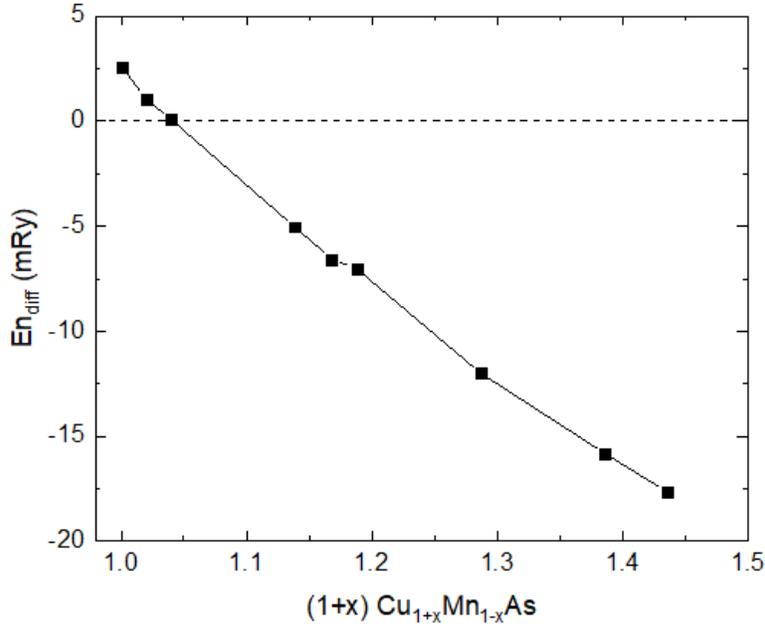

Figure 4. The difference between the total energies of the tetragonal and orthorhombic phases of CuMnAs as a function of extra Cu atom content on Mn sublattice. Positive sign corresponds to the preference for the orthorhombic phase.

*Magnetic properties*

Magnetic susceptibility and differential scanning calorimetry (DSC) were measured on selected polycrystalline samples with various compositions in the temperature range 300-600 K. In Figure 5, data of samples with compositions ($Cu_{1.02(1)}Mn_{0.99(2)}As_{0.99(2)}$, $Cu_{1.15(2)}Mn_{0.87(3)}As_{0.97(3)}$, $Cu_{1.35(2)}Mn_{0.69(3)}As_{0.95(2)}$) are presented. The samples showed antiferromagnetic transitions below the Néel temperatures, magnetization curves at 300 K show linear M(H) behaviour up to 9T (not shown). One exception was a sample with the composition $Cu_{1.02(1)}Mn_{0.99(2)}As_{0.99(2)}$ (Figure 5a) where a ferromagnetic signal appeared below ~315 K. Although one of the previous studies reported ferromagnetism in tetragonal CuMnAs samples [11], we would rather ascribe the ferromagnetic signal to an extrinsic origin (presumably MnAs) as also observed in recently synthetized tetragonal CuMnAs [10] and in some tetragonal CuMnAs thin films when using higher Mn fluxes during the MBE growth [28]. Moreover, the temperature dependence of resistivity, measured on micro-fabricated samples with controlled composition (not shown) from the same batch as the presented sample, does not show any signature of a phase transition around 315 K, confirming the extrinsic nature of this magnetic response.

The Néel temperatures were determined as the inflection point of ($\chi \cdot T$), the values are in agreement with the anomalies in the DSC data. The maximum Néel temperature $T_N$ = 507 K was found for the tetragonal phase with composition closest to the stoichiometric CuMnAs ($Cu_{1.02(1)}Mn_{0.99(2)}As_{0.99(2)}$) being significantly higher than for the orthorhombic CuMnAs. With decreasing Mn content, the Néel temperature decreases monotonically, as shown in Figure 6.

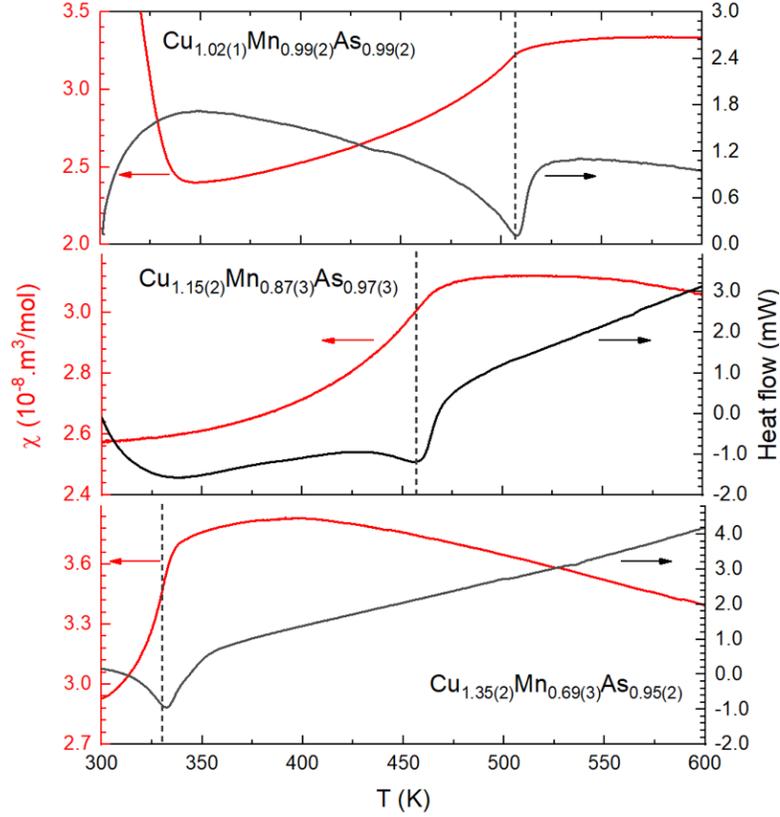

Figure 5. Temperature dependence of Magnetic susceptibility and DSC of samples with compositions: $Cu_{1.02(1)}Mn_{0.99(2)}As_{0.99(2)}$ $Cu_{1.15(2)}Mn_{0.87(3)}As_{0.97(3)}$ and $Cu_{1.35(2)}Mn_{0.69(3)}As_{0.95(2)}$. The short dashed line marks the Néel temperature determined from susceptibility data. The magnetic susceptibility was measured in magnetic field of 1T and calculated as $\chi = M/H$. Magnetization curves are linear up to this field in the whole temperature range.

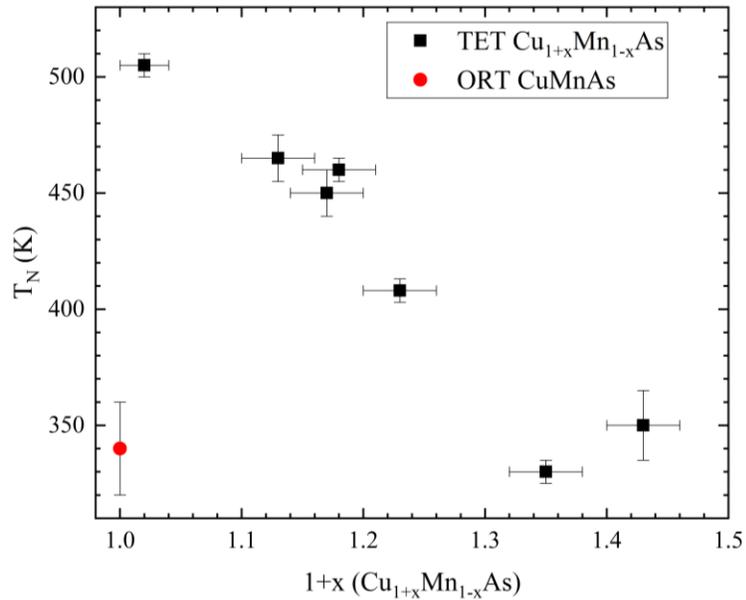

Figure 6. The composition dependence of the Néel temperature in tetragonal CuMnAs samples.

## 4. Conclusions

We have determined the lattice parameters, stable structure and Néel temperatures for $Cu_{1+x}Mn_{1-x}As$ samples up to $x = 0.5$, and small variations of As content (<3%). We have demonstrated that the replacement of Mn by Cu stabilizes the tetragonal phase of $Cu_{1+x}Mn_{1-x}As$, already with as small amounts as $x \sim 0.1$. This stability condition is further modified by possible As excess, whose incorporation into the lattice is, however, more difficult to determine. Therefore its value can only be evaluated with a lower precision.

*Ab initio* calculations show that the difference between the total energy of the tetragonal and orthorhombic phase is very small, and this difference even changes sign with increasing Cu content. Above this threshold, the energy difference monotonously grows in favour of the tetragonal phase, fully supporting the experimental findings.

The highest obtained $T_N$ is 507 K for the tetragonal samples with the closest composition ($Cu_{1.02(1)}Mn_{0.99(2)}As_{0.99(2)}$) to the central point of the ternary system and decreasing with increasing $x$ (decreasing Mn content). The Néel temperatures for the tetragonal samples with $x < 0.2$ are comparable to the temperature of 480 K measured in thin tetragonal CuMnAs films [17] and also in good agreement with previous atomistic spin dynamics simulation [19].

## Acknowledgments

This work was supported by the Czech Science Foundation (Project no. GP14-17102P). Experiments were performed at the Materials Growth and Measurement Laboratory MGML (http://mgml.eu/), which was supported within the program of Czech Research Infrastructures (Project no. LM2011025). E.D.-N. thanks the project financed by ERDF "NanoCent - Nanomaterials centre for advanced applications" (Project No. CZ.02.1.01/0.0/0.0/15_003/0000485). Single crystal XRD measurements were acquired using instruments of the ASTRA laboratory established within the Operation program Prague Competitiveness (Project CZ.2.16/3.1.00/24510).